\documentstyle[eqsecnum,aps]{revtex}
\input epsf
\def\plotone#1{\centering \leavevmode
\epsfxsize= 0.8\columnwidth \epsfbox{#1}}

\def\plotonecita#1{}
\def\plotonecfpa#1{\plotone{#1}}

\def\be{\begin{equation}}
\def\ee{\end{equation}}
\def\bea{\begin{eqnarray}}
\def\eea{\end{eqnarray}}

\def\cmm2{{\,\rm cm^{-2}}}
\def\cm2{{\,{\rm cm}^2}}
\def\cmm3{{\,{\rm cm}^{-3}}}
\def\gcmm3{{\,{\rm g\,cm^{-3}}}}

\def\fun#1#2{\lower3.6pt\vbox{\baselineskip0pt\lineskip.9pt
  \ialign{$\mathsurround=0pt#1\hfil##\hfil$\crcr#2\crcr\sim\crcr}}}

\def\C{{\cal C}}

\def\L{{\cal L}}
\def\muK{\ \mu {\rm K}}
\begin{document}
\twocolumn[\hsize\textwidth\columnwidth\hsize\csname @twocolumnfalse\endcsname
\title{Comparing\\ 
Cosmic Microwave Background Datasets}
\author{L. Knox$^1$, J.~R. Bond$^1$, A.~H. Jaffe$^2$, M. Segal$^1$ and
D. Charbonneau$^1$}
\address{$^1$ Canadian Institute for Theoretical Astrophysics, Toronto, ON M5S 3H8, CANADA}
\address{$^2$ Center for Particle Astrophysics,
  301 LeConte Hall, University of California, Berkeley, CA 94720}
\date{\today}
\maketitle

\begin{abstract}
  To extract reliable cosmic parameters from cosmic microwave
  background datasets, it is essential to show that the data are not
  contaminated by residual non-cosmological signals.  We describe
  general statistical approaches to this problem, with an emphasis on
  the case in which there are two datasets that can be checked for
  consistency.  A first visual step is the Wiener filter mapping from
  one set of data onto the pixel basis of another.  For more
  quantitative analyses we develop and apply both Bayesian and
  frequentist techniques.  We define the ``contamination parameter''
  and advocate the calculation of its probability distribution as a
  means of examining the consistency of two datasets.  The closely
  related ``probability enhancement factor'' is shown to be a useful
  statistic for comparison; it is significantly better than a number
  of $\chi^2$ quantities we consider.  Our methods can be used:
  internally (between different subsets of a dataset) or externally
  (between different experiments); for observing regions that
  completely overlap, partially overlap or overlap not at all; and for
  observing strategies that differ greatly.

  We apply the methods to check the consistency (internal and
  external) of the MSAM92, MSAM94 and Saskatoon Ring datasets.  From
  comparing the two MSAM datasets, we find that the most probable
  level of contamination is 12\%, with no contamination only 1.05
  times less probable, 50\% contamination about 8 times less
  probable and 100\% contamination strongly ruled out at over $2\times
  10^5$ times less probable.  From comparing the 1992 MSAM flight with
  the Saskatoon data we find the most probable level of contamination
  to be 50\%, with no contamination only 1.6 times less probable and
  100\% contamination 13 times less probable.  Our methods can also be
  used to calibrate one experiment off of another.  To achieve the
  best agreement between the Saskatoon and MSAM data we find that the
  MSAM data should be multiplied by (or Saskatoon data divided by):
  $1.06^{+0.22}_{-0.26}$.

\end{abstract}


]

\section{Introduction}

The cosmic microwave background (CMB) is black body radiation with a mean
temperature of $2.728 \pm 0.002$ K \cite{firas}.  This mean is
modulated by a dipole due to our peculiar motion with respect to the
radiation field.  If one removes the dipole, the temperature is
uniform in every direction to $\pm 100 \muK$.  Precision measurement
of these tiny deviations from isotropy can tell us much about the
Universe \cite{forecast}.

Unfortunately, precision measurement of $~100\muK$ fluctuations is not
an easy task.  Even given sufficient detector sensitivity
and observing time, one still has to contend with many possible
contaminants such as side lobe pickup of the $300^\circ$ Kelvin Earth 
and atmospheric noise (even from high-altitude balloons).
In addition there can be contamination of CMB observations by
astrophysical foregrounds.

Despite these difficulties there is good reason to believe that, at
least for some experiments, the signals observed from
sub-orbital platforms are not dominated by contaminants.  One of the
best reasons for believing this comes from the comparisons that have
been done---between FIRS and DMR \cite{Ganga}, Tenerife and DMR
\cite{Lineweaver}, MSAM and Saskatoon \cite{nett95}, two years of Python
data \cite{Ruhl}, and two flights of MSAM \cite{Inman}.  Especially
for the case when data being compared are from two different
instruments, almost the only thing their acquisitions have in common
is that they were observing the same piece of sky--each dataset
has entirely different sources of systematic error.

In addition to confirming the astrophysical origin of the 
estimated signal, comparison can greatly improve the ability
to detect foreground contamination.  Perhaps the best evidence
for the thermal nature of anisotropy comes from the comparison
between the MSAM92 and Saskatoon datasets.  Together, these observations
span a frequency range from 36 GHz to greater than 170 GHz.  In
\cite{nett95} it was found that the spectral index $\beta$ 
($\delta T \propto (\nu/\nu_0)^\beta$) is constrained to be
$\beta = -0.1 \pm 0.2$.  For CMB, free-free and dust over
this frequency range we expect $\beta = 0$, $-1.45$ and $2.25$, respectively.  
The authors conclude that the signals (in the region of overlap) are not 
dominated by contamination from known astrophysical foregrounds, but
are, rather, primarily CMB.

We should not let this apparent success fool us into thinking that
going to the next level of precision will be easy.  There is a big
difference in the level of toleration of contaminants when the goal
switches from detection to precision measurement.  It is likely that
there will be significant levels of contamination (from the
atmosphere, side lobes, and foregrounds) in future sub-orbital
missions.  It may be difficult to convincingly
demonstrate that contamination is low without comparison.

Given the importance of comparison, we feel it is worth improving
upon the methods used previously.  Past treatments
have had to ignore much relevant data, and make uncontrolled
approximations.  This is due to the fact that generally the two
datasets being compared were obtained from instruments observing the
sky in different ways.  The beam patterns and differencing schemes may
differ as in the case of the MSAM/Saskatoon comparison.  In
\cite{nett95} one of the MSAM differencing schemes was approximately
recreated in software in order to do the comparison.  However, no use
of software could change the fact that the MSAM and Saskatoon beam
patterns, although they have fairly similar full-widths at
half-maximum, differ significantly in shape.  Even when the
differencing schemes and beam patterns are the same, there can still
be barriers to a direct comparison.  The two MSAM flights took data
with essentially the same beam pattern and applied the same
differencing, but in this case the direct comparison is frustrated by
the fact that the pixels do not all line up exactly.  Therefore
in \cite{Inman}, pixels within half a beam width of each other
were approximated as being at the same point, and those pixels
with no partner from the other dataset within this distance were ignored.
Half of the data were lost this way.

Here we develop methods of comparing datasets that do not require
any information to be thrown away.  Differences in demodulation
schemes, and effects due to non-overlapping pixels are automatically
taken into account.  The inevitable price we pay for this is
model-dependence.  However, we generally expect the model-dependence
to be small and indeed find it to be so in the case studies shown
here.

An extremely useful tool for visual comparison is the Wiener filter.
Roughly speaking, it allows us to interpolate the
results from one experiment onto the expected results for another
experiment that has observed the sky differently.  After some
notational preliminaries in section II, in section III we 
introduce the Wiener filter in the context of the probability 
distribution of the signal, given the data.  Also in this section
we describe the datasets and apply the Wiener filter to them.

When comparing datasets we are testing the consistency of our model of
the datasets.  We emphasize that meaningful model consistency testing
demands the existence of other models with which to compare.
Therefore we extend our model of the data to include a possible
contaminant and calculate the probability distribution of its
amplitude, given the data.  We find a more limited extension of the
model space to also be useful, in which we only consider one
alternative to no contamination: complete contamination.  We define
the ``probability enhancement factor'' as the logarithm of the ratio
of the probability of no contamination to the probability of complete
contamination.  This Bayesian approach to comparison is described and
applied in section IV.

In section V we discuss and apply frequentist techniques such as $\chi^2$ 
tests.  The probability enhancement factor can also be used
as the basis for a frequentist test---and it is in fact the
well-known likelihood ratio test.  We demonstrate that the
probability enhancement factor has more
discriminatory power than any of the other tests considered.

After a further look at the data with the probability enhancement
factor in section VI, we discuss the fixing of relative calibration
in section VII and possible contamination due to dust in section VII.
Finally we summarize our results in section IX. 

\section{Preliminaries}

Before moving on to a discussion of the various statistics
to be used in comparing datasets, we give some review which
will serve to define our notation, following Ref.\ \cite{pspec}.

In general, CMB observations are reduced to a set of
binned observations of the sky, or pixels, $\Delta_i$,
$i=1\ldots N$
together with a noise covariance matrix, $C_{nii'}$.
We model the observations as contributions from signal and noise,
\be
\Delta_i = s_i+n_i
\ee
We assume that the signal and noise are
independent with zero mean, with correlation matrices given by
\begin{equation}
  C_{Tii'}= \langle s_i s_{i'} \rangle;\quad
  C_{nii'}= \langle n_i n_{i'} \rangle 
\end{equation}
so
\begin{equation}
  \langle \Delta_i \Delta_{i'} \rangle = C_{Tii'}+C_{nii'}
\end{equation}
where $\langle\ldots\rangle$ indicate an ensemble average.
With the further assumption that the data are Gaussian, these
two-point functions are all that is necessary for a complete
statistical description of the data.

One important complication to the above description comes from the
existence of constraints.  Often the data, $\Delta_i$, are susceptible
to some large source of noise, or a not-well-understood source of noise
that contaminates only one mode of the data.  For example, there may be
an unknown offset in the data.
In this case, the
average is usually subtracted from $\Delta_i$.  Similarly, the monopole
and dipole are explicitly subtracted from the all-sky COBE/DMR data, because
the monopole is not determined by the data and the dipole is local in origin.
In general,  placing any constraint on the data or some subset thereof, such as
insisting that its average be zero, results in additional correlations
in $\Delta_i$.  We take this into account by adding these additional
correlations, $C_C$, to the noise matrix to create a ``generalized noise
matrix,'' $C_N$, where $C_N = C_n + C_C$.  In the limit that the
amplitude of $C_C$ gets large, this is equivalent to projecting out
those modes which are now unconstrained by the data, but we find 
this scheme simpler to implement numerically.
Thus in the text below we always
write the noise matrix as $C_N$ instead of $C_n$.  The details of this
procedure for handling the effect of constraints are explained in
\cite{pspec}.

Due to finite angular resolution and switching strategies
designed to minimize contributions from spurious signals
(such as from the atmosphere), the signal is generally
not simply the temperature of the sky in some direction,
$T(\hat x)$, but a linear combination of temperatures:

\begin{equation}
  s_i = \int d^2{\hat x}\; H_i({\hat x}) T({\hat x})
\end{equation}
where $H_i(\hat x)$ is sometimes called the ``beam map'', ``antenna
pattern'' or ``synthesis vector''.  If we discretize the temperature on the sky
then we can write the beam map in matrix form, $s_i = \sum_n H_{in}
T_n$.

The temperature on the sky, like
any scalar field on a sphere, can be decomposed into spherical harmonics
\be 
T(\theta,\phi) = \sum_{\ell m} a_{\ell m}Y_{\ell m}(\theta,\phi).
\ee 
If the anisotropy is {\it statistically}\/ isotropic, {\it i.e.},
there are no special directions in the mean, then the variance of the
multipole moments, $a_{\ell m}$, is independent of $m$ and we can
write: \be \langle a_{\ell m} a^*_{\ell'm'} \rangle = C_\ell
\delta_{\ell\ell'}\delta_{mm'}.  \ee For theories with statistically
isotropic Gaussian initial conditions, the angular power spectrum,
$C_\ell$, is the entire statistical content of the theory in the sense
that any possible predictions of the theory for the temperature of the
microwave sky can be derived from it
\footnote{Non-linear evolution will
produce non-Gaussianity from Gaussian initial conditions
but this is quite sub-dominant for $\ell \lesssim 1000$.}.

The theoretical covariance matrix, $C_{Tii'}$, is related
to the angular power spectrum by
\be\label{eqn:CT}
C_{Tii'} = \sum_\ell {2\ell+1 \over 4\pi} C_\ell W_{ii'}(\ell) \, , 
\ee
where 
\be
\label{eqn:wl}
W_{ii'}(\ell) = \sum_{nn'}H_{in}H_{i'n'}P_\ell(\cos \theta_{nn'})
\ee
is called the window function of the observations and 
$\theta_{nn'}$ is the angular separation between the
points on the sphere labeled by $n$ and $n'$.  

Within the context of a model, the ${\cal C}_\ell$ depend on some
parameters, $a_p$, $p=1\ldots N_p$ which could be the Hubble
constant, baryon density, redshift of reionization, etc.  The
theoretical covariance matrix will depend on these parameters through
its dependence on ${\cal C}_\ell$.  We can now write down the
probability distribution for the data, given the model parameters,
$a_p$:
\begin{eqnarray}
\label{eqn:likelihood}
  P(\Delta|C_T(a_p) I)&=&{1\over(2\pi)^{N/2}
    |C_T(a_p)+C_N|^{1/2}} \times\nonumber\\
 && \exp\left(-{1\over2}
    \Delta^T\left(C_T(a_p)+C_N\right)^{-1}\Delta\right].       
\end{eqnarray}
The $I$ here stands generically for information---in this case the information
that the noise is Gaussian-distributed with zero mean and variance
$C_N$.
\section{Wiener Filters}

\subsection{Derivation}
Bayes' theorem \cite{loredo}
\be
P(s|\Delta I) = P(s|I)P(\Delta|sI)/P(\Delta|I)
\ee
follows from elementary rules of probability.  If we take $P(s|I)$ to
be a Gaussian distribution with zero mean and covariance $C_T$
and $P(\Delta|sI)$ to be a Gaussian with mean $s$ and variance $C_N$
then with a little algebra it follows that 
the probability distribution for the signal,
given the data, $C_T$ and $C_N$, is:
\be
\label{eqn:Pofs}
P(s|\Delta, C_T, C_N) = {\exp\left[-{1\over 2}\left(s-w\Delta\right)^\dagger
M^{-1}\left(s-w\Delta\right) \right]\over
\left[(2\pi)^{N/2}|M|^{1/2}\right]},
\ee
where 
$M \equiv 
\langle \left(s-w\Delta\right)\left(s-w\Delta\right)^\dagger \rangle =
C_T-wC_T$ and 
\begin{equation} 
  \label{eqn:wiener} 
w \equiv C_T\left(C_T+C_N\right)^{-1}  
\end{equation}
is the {\it Wiener filter} \cite{Bunn}.  
As one can immediately see from Eq. (\ref{eqn:Pofs}),
the most probable value of the signal is given by $w\Delta$.  As with
all Gaussian distributions, this most probable value is also the mean:
$\bar s \equiv \int s P(s|\Delta, C_T, C_N) ds = w\Delta$.

Thus the Wiener filter operating on the data provides us with the most
probable estimate of the underlying signal.  Of course, this is the
most probable signal only once we assume a power spectrum, $C_l$,
which is used to calculate $C_T$.  Fortunately this model dependence
is quite weak: the Wiener filter provides a robust estimate of the
underlying signal provided theories are not chosen which are clearly
incompatible with the data.

The Wiener filter can be very helpful for
visualizing the underlying signal.  For example, often the data are
oversampled; that is, there are closely spaced data points with plenty
of scatter and large error bars.  In a sense, the Wiener filter knows
that the high spatial frequency scatter is due to noise and not signal
and performs a smoothing of the data---an interpolation controlled by
the different statistical properties of the noise and signal.

One can also use the dataset to calculate the most probable signal in
some other dataset\footnote{In Ref.\ \cite{Bunn} the Wiener filter
was used to calculate the most probable signal in the Tenerife data,
given the COBE/DMR data.}; let's call the two datasets $\Delta_1$ and
$\Delta_2$, where the subscripts refer here to the entire appropriate
data vector, not the single element at a particular pixel.  Before
getting to $P(s_2 | \Delta_1)$, we describe some notation for joint
datasets.  We represent the total data vector as
\begin{equation}
  \Delta \equiv \left(
    \begin{array}{c}
\Delta_1 \\ \Delta_2
    \end{array}
\right).
\end{equation}
This vector will have a total covariance matrix
\begin{eqnarray}
\label{eqn:jointC}
  \langle \Delta \Delta^\dagger \rangle &=& \left(
    \begin{array}{cc}
      \langle\Delta_1\Delta_1^\dagger\rangle &
      \langle\Delta_1\Delta_2^\dagger\rangle \\
      \langle\Delta_2\Delta_1^\dagger\rangle &
      \langle\Delta_2\Delta_2^\dagger\rangle \\
    \end{array}
  \right)\nonumber\\
  &=&\left(    \begin{array}{cc}
      C_{T11} + C_{N11} & C_{T12}  \\
      C_{T21} & C_{T22} + C_{N22}  \\
    \end{array}
  \right)
\end{eqnarray}
where $C_{Tij}$ represents the theoretical covariance between the
pixels of experiments $i$ and $j$, and $C_{Tij}=C_{Tji}^\dagger$. We
will also define $C_{ij}=C_{Tij} + C_{Nij}$.  We assume that the
experiments have no common noise sources and thus $C_{N12} = 0$.

With this notation established we can now write
\be
\label{eqn:Pofs2}
\begin{array}{l}
P(s_2|\Delta_1, C_T, C_N) \\ 
= {\exp\left(-{1\over 2}
\left(s_2-w_{21}\Delta_1\right)^\dagger
M^{-1}\left(s_2-w_{21}\Delta_1\right) \right)\over
\left[(2\pi)^N |M|\right]^{1/2}} 
\end{array}
\ee
where $M = C_{T22} - w_{21}C_{T12}$, 
\begin{equation} 
w_{21} \equiv C_{T21}\left(C_{T11}+C_{N11}\right)^{-1}
\label{eqn:wiener21}
\end{equation}
and we refer to the Wiener-filtering of dataset one ``onto'' dataset two.

Thus Wiener-filtering provides us with an excellent tool for visual
comparsion of datasets.  Even if each dataset is expressed in different
generalized pixels, since we can Wiener filter one onto the other, we
can compare the signal predictions in the same space.  We will see
applications of this following the next section, which describes the
MSAM and Saskatoon datasets.

The Wiener filter can also be derived without reference to anything
other than the two-point correlation function of the signal and noise.
Assume we want $w$ to be such that the variance $\langle (s -
w\Delta)(s - w\Delta)^\dagger \rangle$ is minimal.  Differentiating
with respect to $w_{ij}$, setting to zero and solving for $w_{ij}$
results in $w = C_T(C_T+C_N)^{-1}$.  Thus the minimum-variance
estimate of the signal does not depend on the Gaussianity of the
signal and noise distributions.  Although, of course, the uncertainty
in the estimate does \cite{Bunn}.

One final expression we will need below is 
the probability distribution for the data itself, $\Delta_2$ (as opposed to
the {\it signal} in the second dataset) given $\Delta_1$ and
relevant matrices.  It is the same as the above after changing
$s_2$ to $\Delta_2$ and $M$ to $M+C_{N22}$.

\subsection{Applications}

For Gaussian signal and noise, the Wiener filter provides the
maximum-Likelihood reconstruction of the signal; it is also optimal in
the minimum-variance sense discussed above.  One can construct a
Wiener filter from the pixelized data space onto the same space or
from the pixelized data space to any other linear combination of map
pixels---such as the map pixels themselves.  Wiener filter maps have
been made for the SK dataset \cite{TegSKmap} and the COBE/DMR dataset
\cite{Bondref1}.  Map-making though is not the most useful means for
comparing observations that are not themselves maps, and it is not
suggested by the statistical techniques we discussed earlier.  Here we
Wiener filter onto the experimental pixel space itself.

\subsubsection{Description of the datasets}
Before jumping into the applications to the Saskatoon and MSAM
datasets we must describe them.  They have considerable spatial
overlap and similar angular resolutions. Otherwise, however, the two
datasets are very different and a comparison provides a strong check
on systematic errors.

\begin{figure}[bthp]
\plotonecita{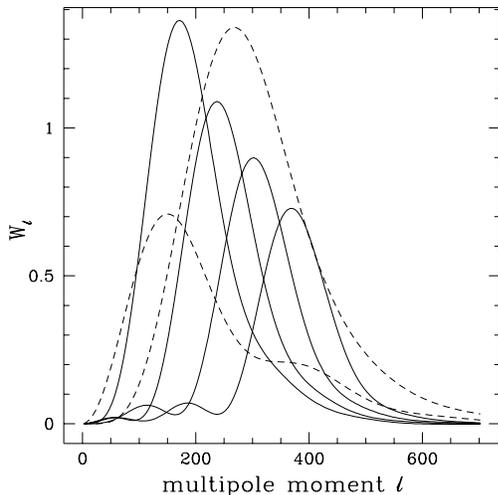}
\plotonecfpa{diagwinds.ps}
\caption[windows]{\baselineskip=10pt 
The diagonal elements of the window function matrix $W_{lij}$
for the four SK ring antenna patterns (solid) and the two MSAM 
antenna patterns (dashed).  These show how the power spectrum contributes
to the variance of the data (see Eq. \ref{eqn:CT}).
}
\label{fig:windows}
\end{figure}

MSAM is a balloon-borne bolometric instrument with approximately
half-degree (fwhm) resolution in 4 frequency bands centered at 170,
280, 500 and 680 GHz \cite{msaminst}.  The data, at each frequency,
are binned into pixels on the sky with two different antenna patterns,
$H$, referred to as 2-beam and 3-beam or single-difference and
double-difference (see corresponding window functions in
Fig. \ref{fig:windows}).  Simultaneously, long time-scale drifts are removed
which has the effect of introducing off-diagonal noise correlations.
From this multi-frequency data, a fit is made to temperature fluctuations
about a 2.73K black-body component and
the optical depth of a dust component.  The dust is assumed to have a
temperature of 20K and emissivity that varies with frequency to the 1.5
power.

The MSAM instrument flew in 1992 \cite{ms1det92}, 1994 \cite{ms1det94} and 
1995 \cite{ms1det95}.  In each year a narrow strip of sky with
nearly constant declination was observed.  The
purpose of the 1994 flight was to confirm the results from the 1992
flight and so each targeted the same strip of sky at $\delta =
82^\circ$ (see Fig. 1).  Note that, due to, for example, imperfect
pointing control, the two flights have slightly different sky
coverage.  The final flight in 1995 observed near declination $\delta
= 80.5^\circ$, chosen to be sufficiently far away from the first two
flights for the signal correlations to be negligible.  Therefore we do
not consider the 1995 flight any further in this paper.

The SK data are reported as complicated chopping patterns ({\it i.e.},
beam patterns, $H$, above) in a circle of radius about $8^\circ$
around the North Celestial Pole. The data were taken over 1993-1995.
Here we only use the 1995 data which were taken with
angular resolution $0.5^\circ$ FWHM at approximately 40~GHz. More
details can be found in \cite{nett95}.

\begin{figure}[bthp]
\plotonecita{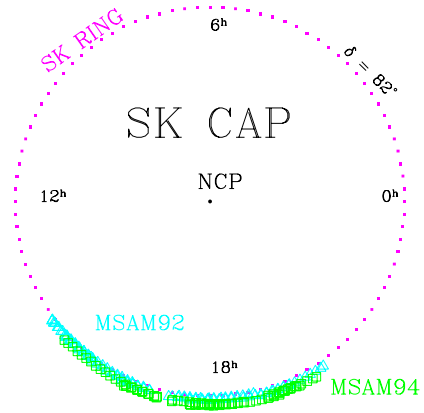}
\plotonecfpa{sky_location.eps}
\caption[windows]{\baselineskip=10pt 
Observation locations.  The SK RING data covered the
entire circle of radius 8 degrees around the NCP.  The
centers of the MSAM92 (MSAM94) pixels are indicated with
triangles (squares).
}
\label{fig:sky_location}
\end{figure}

The bulk of the data were in the ``cap'' configuration:
constant-elevation scans tracing out curved rays from the pole, which
were then binned in RA and subjected to various sinusoidal
demodulation templates in software. Some of the 1995 data ($0.5^\circ$
beam), however, were taken in the ``ring'' configuration, which
isolated the data taken at $\delta = 82^\circ$, put
into 96 RA bins, and then subjected to 3, 4, 5 and 6 point sinusoidal
demodulations, this time along lines of constant declination.  The
ring data window functions are in Fig. \ref{fig:windows}.  The region of
overlap of the SK95 ring data with the MSAM data can be seen in
Fig. \ref{fig:sky_location}.

The calibration of the SK dataset was performed by comparing with the
star Casseiopia A. however, this star's 30--40~GHz flux itself is
poorly determined; hence, the original SK dataset was reported with a
14\% calibration error. More recently, Leitch \cite{leitch} in turn used
the very well-determined amplitude of the CMB dipole itself to
determine the flux of Cas A; this has resulted in a 5\% increase in
the temperature of the SK data (and errors), with a reduced
calibration error of 7\% (the flux of Cas A itself is now determined
to $\sim5\%$, but there are additional sources of
calibration error\cite{nettpriv}). Except for Section \ref{sec:calib},
in the following we do {\em not} include the effects of 
calibration uncertainty.

\subsubsection{Wiener-filtering MSAM92}

\begin{figure}[bthp]
\plotonecita{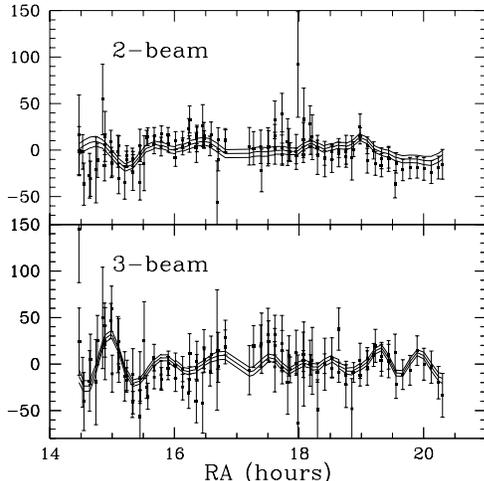}
\plotonecfpa{f92_92.eps}
\caption[f92_92]{\baselineskip=10pt 
An example of Wiener filtering.  The points
with error bars are the MSAM92 pixelized data.
Two-beam in top panel, 3-beam in bottom panel.  
The three curves in each panel
are the Wiener-filtered data bounded by $\pm$ \
one standard deviation.}
\label{fig:f92_92}
\end{figure}

An example of Wiener filtering with Eq.\ (\ref{eqn:wiener}) is shown in
Fig. \ref{fig:f92_92}.  The data points are the values of the pixelized
data, located horizontally according to the right ascension of the
center of the pixel.  The dependence of the pixels on declination and
twist has been suppressed.  The error bars are from the diagonal part of
the (non-diagonal) noise covariance matrix.  The central curve is the
Wiener-filtered data and the bounding curves indicate the 68\%
confidence region for the signal.  Because of the difference between the
noise covariance matrix and the signal matrix, the Wiener filter
essentially assumes that the high frequency behavior is noise and
therefore smooths out the data.  This smoothing is complicated by the
off-diagonal noise correlations which explains some apparent
disagreements between the data and the Wiener-filtered data.  For
example, around 20 hours in the top panel, the Wiener-filtered data are
consistently above a number of the data points.

\begin{figure}[bthp]
\plotonecita{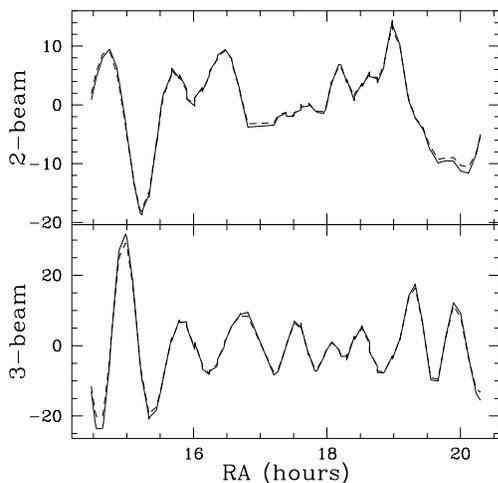}
\plotonecfpa{f92_92,92flat.ps}
\caption[f92_92,92flat]{\baselineskip=10pt 
Wiener filter model-dependence for MSAM92.
The standard CDM (flat) spectrum was assumed for the solid (dashed)
curve.}
\label{fig:f92_9292flat}
\end{figure}

The Wiener filter is model-dependent---one must know (or assume)
covariance matrices for the noise and signal.  Presumably the noise
covariance matrix is well-known and so the model-dependence resides in
the choice of angular power spectrum.  Of course, we can gain some
knowledge of the angular power spectrum by performing a likelihood
analysis of the data.  The Wiener filter is generally quite robust to
changes in the angular power spectrum that are smaller than those
that significantly alter the likelihood---even large changes usually
have very little effect.  We demonstrate this robustness here with
Fig. \ref{fig:f92_9292flat} which shows the Wiener-filtered data for a
standard CDM spectral shape and also for a flat spectrum ($\C_\ell =
$\ constant).

\subsection{Wiener-filtering MSAM94 onto MSAM92}

\begin{figure}[bthp]
\plotonecita{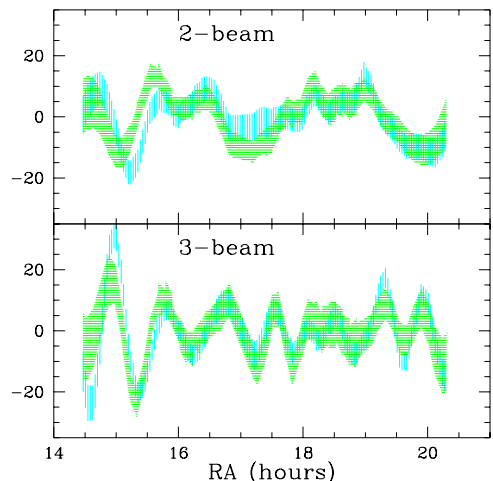}
\plotonecfpa{f92_shade9294.eps}
\caption[f92_shade9294_real94]{\baselineskip=10pt 
Wiener filters onto 1992 pixels for 1992 data (vertical
lines) and 1994 data (horizontal lines).  The curves
are realizations consistent with the 1994 data.
Two-beam in top panel, three-beam in 
bottom panel.}
\label{fig:f92_shade9294}
\end{figure}

Besides Wiener filtering the data onto its own pixel space, we can
Wiener filter it onto another pixel space (Eq.\ \ref{eqn:wiener21}).
This provides an excellent visual tool for checking compatibility of
results.  We show this first for the Wiener filtering of MSAM94 onto
MSAM92, together with MSAM92 onto MSAM92 from the previous subsection.
Notice that in Fig.  \ref{fig:f92_shade9294} the 68\% confidence
regions mostly overlap each other.  One can see the MSAM94 region get
wider at either extreme in RA.  This is because the MSAM94 pixels have
a slightly shorter RA extent than the MSAM92 pixels ($14.9^{\rm h}$ to
$20.1^{\rm h}$ compared to $14.5^{\rm h}$ to $20.3^{\rm h}$).

One can see in the figure that many features are seen by both
datasets; they agree quite well.  The most significant differences
between the two estimates of the signal are in the region of 15.5
hours for the 2-beam signal and 14.5 hours for the 3-beam
signal.  We will discuss these slight anomalies later.

\subsection{SK95 onto MSAM92 and MSAM92 onto SK95}

\begin{figure}[bthp]
\plotonecita{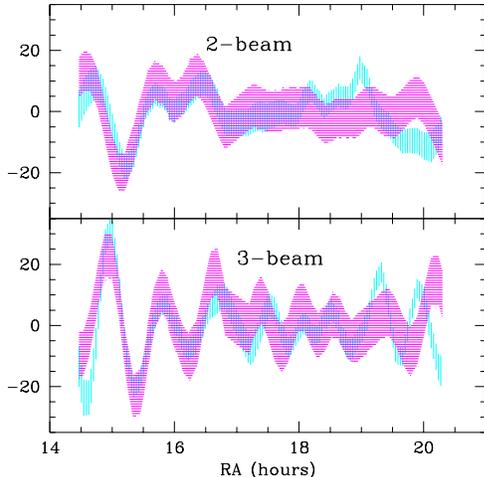}
\plotonecfpa{f92_shade9295.eps}
\caption[f92_shade9295]{\baselineskip=10pt 
The 1992 (vertical lines) and 1995 data (horizontal lines) 
Wiener-filtered onto the 1992 pixels.
}
\label{fig:f92_shade9295}
\end{figure}

Figure \ref{fig:f92_shade9295} shows the same thing as Fig.
\ref{fig:f92_shade9294} except that MSAM94 has been replaced with
Ring95.  Once again, the first impression is of general agreement,
although the discrepancies here (at large RA) appear to be more
significant than those seen in the MSAM92/MSAM94 comparison.

\begin{figure}[bthp]
\plotonecita{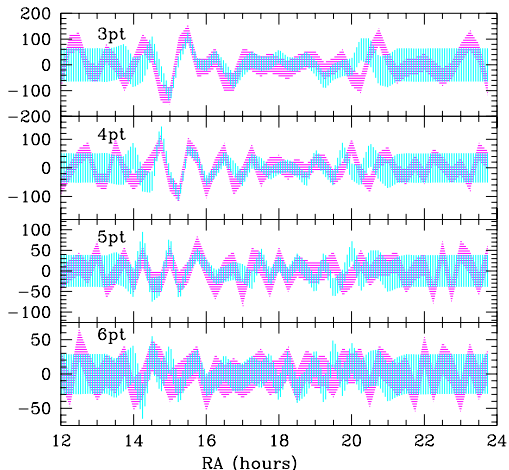}
\plotonecfpa{f95_shade9592.eps}
\caption[f95_shade9592]{\baselineskip=10pt 
Wiener filters onto 1995 pixels for 1992 data (vertical
lines) and SK 1995 data (horizontal lines).}
\label{fig:f95_shade9592}
\end{figure}

We can also filter the MSAM92 data onto the four Ring templates, as
shown in Fig. \ref{fig:f95_shade9592}.  We have chosen the range of
this plot to extend in RA further than the MSAM coverage.  This allows
one to see how the constraint behaves outside of the region of MSAM's
influence.  Notice that the errors don't become infinite.  This is
because of the prior information that went into the estimate of the
probability distribution, i.e., the assumed power spectrum.  Also note
that the data have some influence slightly beyond the limit of the sky
coverage.  The dominant reason for this is the spatial extent of the
antenna patterns.  In addition, the intrinsic correlations (assumed in
the prior) extend the influence to slightly beyond where the antenna
response is zero.

With two dimensional Wiener filter maps (as with
any two dimensional map) it is difficult
to plot both the map and a confidence region
expressing the level of uncertainty as we have
done here for essentially one dimensional data.
In 2D it is therefore often useful to show,
in addition to the mean signal, several realizations
consistent with its probability distribution (Eq. \ref{eqn:Pofs} or
\ref{eqn:Pofs2}).  Looking at several realizations allows
one to see which features are significant and which aren't.
Realizations can also be useful in the 1D case to make
up for the fact that the confidence region does not
contain any information about correlated uncertainties.
For the applications here, though, we have not found them
to be useful and so have not shown any.  

\section{Bayesian Comparison}

A natural question to ask is, ``How consistent are the two
datasets?''.  The Wiener filter gives a visual, qualitiative answer to
the question, but we would also like some quantitative answers as
well.  A better-formulated question is, ``Is my model of the data an
adequate description of the two datasets together?''.  To answer this
question, one can extend the model of the data to include a residual
and then check to see if this extension increases the likelihood.  For
example, one could add a residual that is Gaussian-distributed with
zero mean: 
\bea 
\Delta_i & = & s_i + n_i + r_i\nonumber \\ \langle
\Delta_i \Delta_j \rangle & = & C_{T,ij} + C_{N,ij} + C_{{\rm
res},ij}.  
\eea 
Further restrictions on the form of $C_{{\rm res},ij}$
must be made for the problem to not be degenerate.  One could choose
$C_{{\rm res},ij}$ to be appropriate for a particular foreground
contaminant \cite{Jaffedust}, increased noise \cite{Bondref1}, or
anything else that inspection of the data, combined with prior
knowledge, has led the analyzer to suspect.  Below we describe a
particular choice of $C_{{\rm res},ij}$ that is useful in the absence
of any hints as to the likely nature of a possible contaminant.

\subsection{The contamination parameter, $\gamma$}

To test the consistency of the pairs of datasets -- or
rather, to test the adequacy of our model of the datasets --
we introduce the following residual:
\be
\Delta_1 = s_1+n_1+\gamma_1 r_1
\ee
and likewise for $\Delta_2$.  To reduce the number of parameters
in this model for the residual, we set $\gamma = \gamma_1 = \gamma_2$.
Now we must specify the probability distribution of $r$.  For simplicity,
let's take it to be a Gaussian random variable with zero mean.
Clearly we want the cross-term in the variance to be zero 
($\langle r_1 r_2^\dagger \rangle = 0$) since we have in mind contaminants 
that are particular to each dataset.  There is a lot of freedom
in the choice of $\langle r_1 r_1^\dagger\rangle$ and 
$\langle r_2 r_2^\dagger\rangle$---once 
again for simplicity let's take these to be equal to $C_{T11}$ and
$C_{T22}$.  

We have just added one parameter to whatever other parameters
we were using to define the power spectrum.  The model for the
power spectrum we use here is standard CDM, with the amplitude
as the only free parameter.
We have expressed the amplitude as $\sigma_8$---the  {\it rms}
fluctuations in mass in 8h$^{-1}$\ Mpc spheres.  
The experiments in question do not have sufficient dynamic
range to {\it strongly} constrain more than this one parameter.
For COBE-normalized standard CDM, $\sigma_8=1.2$.

We can now explicitly show the complete parameter dependence of 
the covariance matrix in our model, by modifying 
Eq. (\ref{eqn:jointC}) to:
\begin{eqnarray}\label{eqn:Cgamma}
C =&\left(    \begin{array}{cc}
      \sigma_8^2\left(1+\gamma^2\right)\tilde C_{T11} + C_{N11} 
   & \sigma_8^2 \tilde C_{T12}  \\
      \sigma_8^2 \tilde C_{T21} 
   &\sigma_8^2\left(1+\gamma^2\right) \tilde C_{T22} + C_{N22}  \\
    \end{array}
  \right)
\end{eqnarray}
where the tilde means the quantity is evaluated for $\sigma_8=1$.  

\begin{figure}[bthp]
\plotonecita{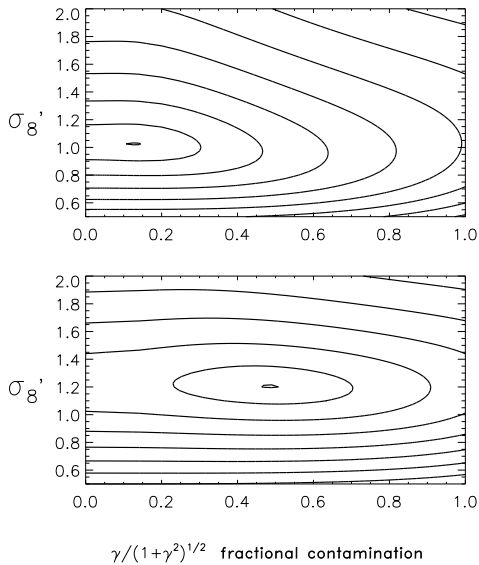}
\plotonecfpa{gamma_9294_9295.eps}
\caption[gamma92949295]{\baselineskip=10pt 
Contours of the likelihood of $\sigma^\prime_8$ vs. the fractional
contamination for the MSAM92 and MSAM94 datasets (top panel)
and the MSAM92 and SK95 datasets (bottom panel).  The contours
indicate reductions in probability from the maximum by factors
of $e^{1^2/2}$, $e^{2^2/2}$,$e^{3^2/2}$, etc.
}
\label{fig:gamma}
\end{figure}

We prefer to work with a slightly different parameterization (spanning
the same model space) by replacing $\sigma_8^2$ with  
$(\sigma_8^\prime)^2 \equiv \sigma_8^2 (1+\gamma^2)$ which is
the amplitude for the variance of the signal and the contaminant combined.
We prefer $\sigma_8^\prime$ to $\sigma_8$ since its probability
distribution of this quantity should be relatively independent of
the level of contamination.  Further, we prefer to use the fraction
of contamination, $\gamma/\sqrt{(1+\gamma^2)}$ rather than the contamination
parameter itself.
Probability distributions for
$\sigma_8^\prime$ and $\gamma/\sqrt{(1+\gamma^2)}$ can be seen in Fig.
\ref{fig:gamma}.

One can see from the shape of the contour curves that
$\gamma/\sqrt{(1+\gamma^2)}$ and $\sigma_8^\prime$ are very nearly uncorrelated.
The reason is that the dominant contribution to the determination
of $\sigma_8$ comes from terms in the likelihood proportional
to $\Delta_i \Delta_j$ where $\Delta_i$ and $\Delta_j$ are in
the same dataset, whereas $\gamma$ is entirely determined by
the cross-terms.  

The most probable level of contamination indicated
by the MSAM92/MSAM94 comparison is about 12\%.  However,
there is virtually no evidence for non-zero contamination
since the probability of zero contamination is only about
5\% less.   
Complete contamination is strongly
ruled out at more than $\exp(5^2/2) \simeq 2.7\times 10^5$ times less
probable.  The MSAM92/SK95 datasets are much less constraining on the amount
of contamination that may be present.  While 50\% 
is the most probable value, total contamination and no contamination
are only about 13 and 1.6 times less likely respectively.  

The residual, as we have modeled it here, is a particularly difficult
one to constrain since it very nearly has the same statistical properties
as the signal.  We note that this is desirable in the sense that the
ability to constrain the residual comes entirely from the comparison --
that is, each dataset, by itself, has no constraint on the fractional
contamination.  Thus this model for the residual is a strong test
of the agreement {\it between} the two datasets, rather than anything
internal to them.

\subsection{The probability enhancement factor, $\beta$}

For many purposes, a much smaller extension into
alternative hypothesis space may be useful.  In
particular, instead of examining a continuum, one
could just compare the model with $\gamma=0$ to
the model with $\gamma = \infty$, at fixed $\sigma_8^\prime$.
The interesting
quantity is how much more probable one model is
than the other, a quantity referred to as the odds.
This particular odds, or rather its logarithm, we
refer to as $\beta$ and call it 
the probability enhancement factor:
\begin{equation}
\label{eqn:betadef}
  \beta\equiv\ln{P(\Delta_1\Delta_2|H_0)\over
    P(\Delta_1 \Delta_2|H_\infty)}
\end{equation}
where $H_0$ (not to be confused with the present value
of the Hubble constant!) is the hypothesis 
that $\gamma=0$ and $H_\infty$ is the hypothesis that $\gamma=\infty$.
Both hypotheses are understood to be fixed at the same $\sigma_8^\prime$.
One can see from Eq. \ref{eqn:Cgamma} that the cross-terms connecting
the two different datasets in the covariance matrix $C$ vanish when
$\gamma \rightarrow \infty$ with $\sigma_8^\prime$ fixed.  Therefore
we can also write $\beta$ as  
\begin{equation}
\label{eqn:betadef2}
  \beta=\ln{P(\Delta_1\Delta_2|C)\over
    P(\Delta_1|C)P(\Delta_2|C)}=\ln{P(\Delta_1|\Delta_2,C)\over
    P(\Delta_1|C)}
\end{equation}
where $C$ is understood to be $C$ in Eq. \ref{eqn:Cgamma} with
$\gamma=0$ and the second equality follows from the use of
$P(AB|C)=P(A|BC)P(B|C)$.  This second equality gives rise to another
interpretation of $\beta$:
$\beta$ indicates how much
more probable dataset 1 is given that dataset 2 exists than it would
be without the existence of dataset 2.  And by the symmetry of the
definition of $\beta$ we know that the statement is true under
switching of 1 and 2.
\footnote{There is even a third interpretation of 
$\beta$ as the log of the ratio of probability of no relative
pointing error, to that of a gross relative pointing error which leaves the
fields completely uncorrelated.}

\begin{figure}[bthp]
\plotonecita{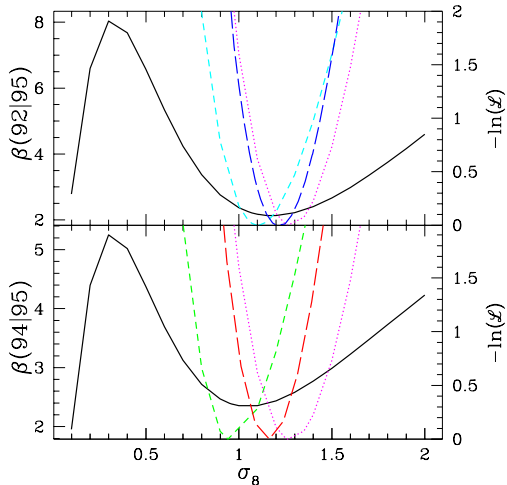}
\plotonecfpa{2betas_lek.eps}
\caption[2betas]{\baselineskip=10pt 
Probability enhancement factor $\beta(92,95)$ (top panel) 
and $\beta(94,95)$ (bottom panel) as a function of $\sigma_8$ (solid curves).
Also shown are $-\delta \ln {\cal L}$ for individual and joint
datsets.  Identifying these curves by their minima, they are,
from left to right: MSAM92, MSAM92+SK, SK in the top panel and
MSAM94, MSAM94+SK, SK in the bottom panel.}
\label{fig:2betas}
\end{figure}

The probability enhancement factor, like the Wiener filter, depends on
the assumed power spectrum used to calculate the theoretical
covariance matrices.  We find that for our parametrized model, within
the most likely region of parameter space, the dependence of $\beta$
on the parameter is weak.  In Fig. \ref{fig:2betas} we see the
dependence of $\beta(92,95)$ and $\beta(94,95)$ on
$\sigma_8$.\footnote{To be precise, we mean $\sigma_8^\prime$ but in
the following we drop this prime for simplicity and also because keeping
the prime does not make sense in the context of the interpretation of
$\beta$ as the increase in probability of one dataset given the other
dataset.}  Notice that it doesn't make much difference to $\beta$
whether one uses the best fit value of $\sigma_8$ given by one of the
two experiments or by the joint likelihood---or indeed by anything in
the 68\% confidence region because the dependence of $\beta$ on
$\sigma_8$ is quite flat in this region.

As can be seen from the log likelihood curves, the different datasets
prefer slightly different values of $\sigma_8$\footnote{Some of this
discrepancy may be due to calibration uncertainty which is not
included in these log likelihood curves.  We address this issue in a
later section.}.  For all calculations of $\beta$ below and for the
Wiener filtering in the previous section we have chosen a value of
$\sigma_8=1.2$, in between the preferred values for SK and MSAM. It is
also the normalization for this power spectrum suggested by the DMR
data.  

\section{Frequentist Statistics}

We now discuss $\beta$
from the frequentist perspective.  The frequentist approach to
checking the consistency of a dataset is to invent some function of
the data, called a statistic, and then to compare the measured value
of the statistic to its probability distribution under various
hypotheses.  The probability enhancement factor, $\beta$, can be
viewed as a statistic since it is a function of the data.  In fact, it
is the logarithm of the well-known likelihood ratio statistic---in
this case the ratio of the likelihood of $H_0$ to the likelihood of
$H_\infty$.

Some statistics are better than others at distinguishing among competing
hypotheses.  In this section, we see how $\beta$ and 
other statistics fare at discriminating between hypotheses $H_0$ and
$H_\infty$.

\subsection{Probability distributions of quadratic statistics}
We restrict ourselves to studying quadratic functions of the
data, for which we have analytic expressions for the mean and
variance.  In addition to various different $\chi^2$ quantities
(to be defined below), the probability enhancement factor---
due to the logarithm in the definition---is also a quadratic
function of the data:
\bea
\beta& =& (N/2)\ln |C| + {1\over 2} \Delta^T C^{-1} \Delta\nonumber\\
&-&(N_1/2)\ln |C_{11}| -{1\over 2}\Delta_1^T C_{11}^{-1} \Delta_1\nonumber\\
&-&(N_2/2)\ln |C_{22}| -{1\over 2}\Delta_2^T C_{22}^{-1} \Delta_2.
\eea
which follows from Eq. \ref{eqn:betadef2}.  Since it is a quadratic
function of the data, it is straightforward to calculate the mean
and variance.

In general, any quadratic function of the data, $Q \equiv \Delta^\dagger M
\Delta + {\rm constant}$, has a mean under hypothesis $X$ of 
\be
\label{eqn:mean}
\bar Q_X \equiv \langle Q \rangle_X = {\rm Tr}\left(C_X M\right)
+ {\rm constant}
\ee
and a variance of 
\be
\label{eqn:var}
\delta Q_X^2 \equiv \langle \left(\bar Q_X - Q\right)^2\rangle_X =
2{\rm Tr}\left(C_X M C_X M\right)
\ee 
where hypothesis $X$ is specified by $C_X \equiv \langle \Delta 
\Delta^\dagger \rangle_X$.

For the case of $\beta$ we have, for hypotheses $H_0$ and $H_\infty$:

\bea
\label{eqn:betastats}
\langle \beta \rangle_0 &=& {1\over 2}
\ln\left({|C_{11}|^{N_1}|C_{22}|^{N_2}\over |C|^N}\right) \nonumber \\
\langle \left(\beta - \langle \beta \rangle_0\right)^2\rangle_0 &=&
{\rm Tr}\left(w_{12}w_{21}\right) \nonumber \\
\langle \beta \rangle_\infty &=& \langle \beta \rangle_0+{1\over 2}
{\rm Tr}\left(1-C_\infty C^{-1}\right)\nonumber \\
\langle \left(\beta - \langle \beta \rangle_\infty\right)^2\rangle_\infty &=&
{1\over 2}
{\rm Tr}\left[\left(1-C_\infty C^{-1}\right)\left(1-C_\infty C^{-1}\right)\right]
\eea
Note that if the experiments have nothing to do with each other
($C_{12} = 0$) then the numerator and denominator of the
argument of the logarithm are equal and therefore 
$\langle \beta \rangle_0 = 0$
as we expect from the definition of $\beta$ in Eq. \ref{eqn:betadef}.

\begin{figure}[bthp]
\plotonecita{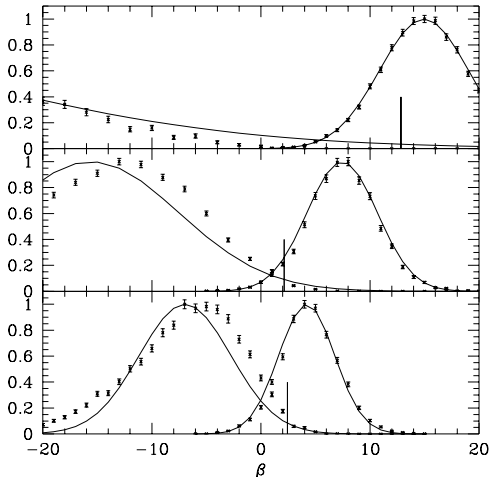}
\plotonecfpa{P_of_beta.eps}
\caption[P_of_beta]{\baselineskip=10pt 
The measured values of $\beta$ (vertical lines) 
and its (arbitrarily  normalized) probability distribution functions 
under the two hypotheses.
From top to bottom: $\beta(92,94)$, $\beta(92,95)$ and
$\beta(94,95)$.  The curves peaking at positive $\beta$ are 
estimates of $P(\beta|H_0)$ and those peaking at negative $\beta$ are
estimates of $P(\beta|H_\infty)$.  The points with error bars are the results
of a Monte Carlo calculation while the solid curves
are Gaussians with the analytically calculated means and variances.}
\label{fig:P_of_beta}
\end{figure}

Given the observed value of $\beta$, we can assess the validity of the
two hypotheses by calculating the probability distribution of $\beta$
under each hypothesis.  As shown above we can calculate the mean and
variance analytically.  
To calculate the entire (non-Gaussian) distribution function
though, we have used the Monte Carlo method.  The results are plotted
in Fig. \ref{fig:P_of_beta} for the three possible pairings of the
three datasets.  The Monte Carlo method is quick because we first
rotate to a basis where everything is diagonal and then make the
realizations.  The rotation to the diagonal basis only needs to be
found once.  The plots shown use between 4000 and 17000 realizations.
Notice that the distribution of $\beta$ under $H_0$ is
well-approximated by a Gaussian.  The deviations from Gaussianity are
larger under $H_\infty$.

We see in the figure that $\beta(92,94)=13$ which is consistent with
the expected range for hypothesis 0 of $15 \pm 4.1$.  As a
measure of the consistency, we have calculated the probability of
getting a $\beta$ greater than this to be 0.70.  We also see that
under hypothesis $H_\infty$ such a value of $\beta$ is extremely
unlikely; the probability of getting a $\beta$ greater than the
measured one is less than 1\%.  We also find consistency with $H_0$
for the other two pairs of datasets: $\beta(92,95) = 2.1$
(c.f. $\langle \beta \rangle_0 = 7.4\pm 3.2$) and $\beta(94,95) =
2.4$ (c.f. $\langle \beta \rangle_0 = 4.4\pm 2.6$).  For both of
these, under hypothesis $H\infty$, the probability of getting a value
of $\beta$ as high or higher than the measured one is $1\%$.  

\subsection{Comparison of comparisons}
There are a handful of other quadratic functions of the data
one might consider using for comparison of datasets.  Here we
define the ones under consideration by specifying the data vectors
on which they are based:
\bea
\chi^2_J: & & \Delta \\
\chi^2_w: & & \Delta - w\Delta \\
\chi^2_{w1}: & &(\Delta_1 - w_{12} \Delta_2) \\
\chi^2_{w2}: & &(\Delta_2 - w_{21} \Delta_1) \\ 
\chi^2_{w12}: & &(w_{12}\Delta_2 - w_{11} \Delta_1) \\
\chi^2_{w21}: & &(w_{21}\Delta_1 - w_{22} \Delta_2) 
\eea
We clarify what we mean by two examples: 
\be
\chi^2_J = \Delta^\dagger M^{-1} \Delta
\ee
where $M = \langle \Delta \Delta^\dagger \rangle_0 = C$, and
\be
\chi^2_{w12} = (w_{12}\Delta_2 - w_{11}\Delta_1)^\dagger M^{-1}
(w_{12}\Delta_2 - w_{11} \Delta_1),
\ee
where 
\bea
M &\equiv &\langle \left(w_{12}\Delta_2 - w_{11} \Delta_1\right)
\left(w_{12}\Delta_2 - w_{11} \Delta_1 \right)^\dagger\rangle_0 
\nonumber\\
&=& \left(w_{11} - w_{12}w_{21}\right)C_{T11} + 
\left(w_{12}-w_{11}w_{12}\right) C_{T21}.
\eea
The $J$ stands for joint, since this is the $\chi^2$ quantity
in the joint likelihood function, $P(\Delta|C)$.  
It is straightforward to show that $\chi^2_J = \chi^2_w$, but, other
than this relationship, the above $\chi^2$s are all independent quantities.

\begin{figure}[bthp]
\plotonecita{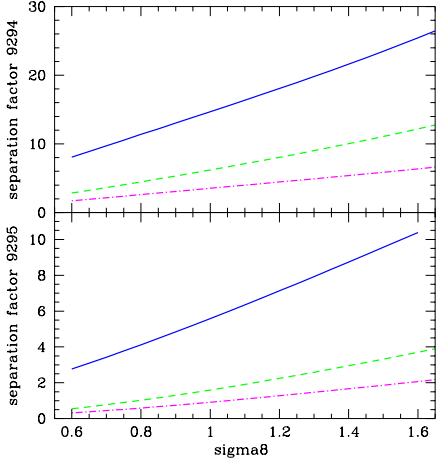}
\plotonecfpa{sepfac_9294_9295.eps}
\caption[sepfac]{\baselineskip=10pt 
Separation factors for $\beta$ (blue, solid), $\chi^2_{w12}$
(green, dashed) and $\chi^2_J$ (magenta, dot-dashed). The
top panel is for the 92/94 comparison and the bottom panel
for the 92/95 comparison.  For $\chi^2_{w12}$ the smaller
dataset is taken to be dataset 1.
}
\label{fig:sepfac}
\end{figure}

To judge the discriminating power of all our quadratic statistics,
we use the {\it separation factor},
\be
|Q_0-Q_\infty|/\delta Q_0, 
\ee
where $Q_0$, $Q_\infty$ are the means under the two hypotheses and 
$\delta Q_0$ is the standard error under $H_0$.
The separation factor is shown as a function of $\sigma_8$ in
Fig.\ \ref{fig:sepfac}.  To avoid clutter, only two of the $\chi^2$
quantities are shown, $\chi^2_J$ and $\chi^2_{w12}$.  The separation
factors for the other $\chi^2$s are bounded by these two.  

One can clearly see the superiority, under
this measure, of the Bayesian-motivated probability enhancement factor.
For example, for $\sigma_8=0.6$, if we assume $H_0$,
it requires an $8\sigma$ fluctuation to get $\beta = \langle \beta
\rangle_\infty$ but only a $2\sigma$ ($3\sigma$) fluctuation to get
$\chi^2_J = \langle \chi^2_J \rangle_\infty$ 
($\chi^2_{w12} = \langle \chi^2_{w12} \rangle_\infty$).  The increase
in all the separation factors with increasing $\sigma_8$ is expected since
discriminating power should increase with increasing signal-to-noise
of the measurements.

The separation factor, as we have defined it, 
is the separation between the expected value of the two hypotheses
in units of the standard error assuming $H_0$ ($\delta Q_0$).
One might also choose as another measure of discriminating power,
this separation in units of the standard error assuming $H_\infty$
($\delta Q_\infty$).  In showing that $\beta$ performs well
under this measure we are assisted by a theorem:  the
likelihood ratio test is {\it most powerful}.

A simple hypothesis test can be made from any statistic by choosing
some critical value $Q^*$: if $Q > Q^*$ then reject $H_0$; otherwise, accept
$H_0$\footnote{This assumes $Q_0 < Q_\infty$, if not then the test should
be changed so that $H_0$ is rejected when $Q > Q^*$.}.  
Statisticians discuss the {\it size} and {\it power} of a test
designed to discriminate between two hypotheses.  The size of the test
is the probability of rejecting $H_0$ if $H_0$ is true while the power
is the probability of rejecting $H_0$ if $H_\infty$ is true.  Clearly, we
want the test to be such that the size is small and the power is
large.  By changing $Q^*$ we can choose the size.  The test based
on the likelihood ratio statistic has the property that, for a given
size, it is most powerful---that is, no other test with the same size
has a greater power.  For a discussion of the likelihood ratio
statistic in the context of CMB observations see, e.g.,
\cite{Readheadetal}.

To see the relevance with our separation factor, let's set 
$Q^* = Q_0$.  Let's further assume that we are in the asymptotic
limit of large datasets so that all probability distributions
are Gaussian.  With $Q^* = Q_0$, the size of the test will be 0.5
for all statistics.  Since the size of this test is the same
for all statistics, we know that the likelihood ratio
test ($Q = \beta$) will have the largest power.  For $Q^* = Q_0$ the
power is given by
\bea
{\rm power} &=&1/2+{1\over \sqrt{ 2\pi}\delta Q_\infty}\int_{Q_0}^{Q_\infty}
\exp\left(Q-Q_\infty\right)^2/\left(2\delta Q_\infty^2\right) \nonumber \\
&=&1/2+{\rm erf}\left(\left(Q_0-Q_\infty\right)/\sqrt{2\delta Q_\infty^2}\right)/2.
\eea
Since the error function monotonically
increases with its argument, we see that the separation between $Q_0$ 
and $Q_\infty$ in units of $\delta Q_\infty$ will
always be largest for the likelihood ratio statistic, $\beta$.

We end this section with a brief consideration of one more $\chi^2$
quantity.  One could ask
if there is a set of map pixels, $T$, that is consistent with 
the noise distribution:
\be
\chi^2_n \equiv (\Delta - s)^\dagger C_n^{-1}(\Delta -s) \ \ ; \ \ s \equiv HT
\ee
Because of its model independence, one might also think that
$\chi^2_n$ is a compelling choice for testing the consistency of two
datasets.  However, if the pixels for the two datasets are slightly
different, then it will almost certainly be the case that a set of map
pixels can be found that gives a reduced $\chi^2_n$ near unity.  The
problem is that this sky map may contain sharp spikes, highly
inconsistent with our prior assumptions.

\section{Applying $\beta$ to subsets of data}
We have also calculated $\beta$ for various pairings of subsets of the
data; the results are in the Table.  All but one pairing (to be
discussed later) have values of $\beta$ within $2\sigma$ of $\langle
\beta \rangle_0$.  Note that the last 8 rows of the table are the
results for internal consistency checks.  

Also included in the table are the values of  
$\chi^2_{w12}$.  Under the separation factor criterion, this
was the best other quadratic statistic.  It is also
of particular relevance to   
Figures \ref{fig:f92_shade9294}, \ref{fig:f92_shade9295} 
and \ref{fig:f95_shade9592} since these show
the data vectors on which $\chi^2_{w12}$ is based.

Most of the reduced $\chi^2_{w12}$ values are comfortably close to unity.  
The probability of exceeding $\chi^2$ is less than $5\%$ for
only one of the entries---the 95\_4,95\_5 internal consistency
check for which the probability is less than 1\%.
\vfill
\begin{table}
\begin{center}
\begin{tabular}{|c|c|c|c|c|c|} \hline

datasets & $\beta$ & $\langle\beta\rangle_0 \pm \delta \beta_0$ & 
$\langle\beta\rangle_\infty \pm \delta \beta_\infty$ & 
$\chi_{w12}^2/\nu$ & $\nu$\\ \hline

92,94\_2 & 5.7 & $10.9 \pm 3.6$ & $-37.6 \pm 20.0$ & 1.08  & 218\\
92,94\_3 & 14.5 & $11.2\pm3.6$ & $-39.0 \pm 20.1$ & 1.05 & 218\\
92,94 & 12.8 & $15.0\pm 4.1$ & $-58.4 \pm 27.4$ & 1.02 & 218\\ \hline

92,95\_3 & -2.5 & $4.4 \pm 2.5$ & $-8.5 \pm 5.4$ & 1.12 & 218\\
92,95\_4 & 4.6 & $3.2\pm 2.2$ & $-5.3 \pm 3.6$ & 1.11  & 218\\
92,95\_5 & -1.2 & $1.6 \pm 1.7$ & $-2.031 \pm 1.4$ & 1.05 & 218\\
92,95\_6 & -0.29 & $0.56 \pm 1.03$ & $-0.61 \pm  0.49$ & 1.06 &218 \\ 
92,95 & 2.13 & $7.4\pm 3.2$ & $-15.6 \pm 8.1$ & 1.15  & 218\\ \hline
94,95\_3 & 2.6 & $2.71 \pm 2.08$ & $-4.27 \pm 3.05$ & 1.02 & 170\\
94,95\_4 & 1.4 & $1.94 \pm 1.82$ & $-2.69 \pm 2.01$ & 1.05 & 170\\
94,95\_5 & -0.31 & $0.99 \pm 1.35$ & $-1.14 \pm 0.85$ & 1.06 & 170\\
94,95\_6 & -0.99 & $0.35 \pm 0.82$ & $-0.365 \pm 0.29$ & 0.96 & 170\\ 
94,95 & 2.437 & $4.4 \pm 2.63$ & $-7.29 \pm 4.2$ & 1.05 & 170\\ \hline

92\_2,92\_3 & 8.29 & $8.80 \pm 3.185$ & $-31.6 \pm 18.775$ & 1.16 & 109\\ 
94\_2,94\_3 & 6.81 & $11.1 \pm 3.418$ & $-52.4 \pm 30.4$ & 0.93 & 85\\ 
95\_3,95\_4 & 1.72 & $5.2 \pm 2.99$ & $-7.12 \pm 3.12$ & 1.25 & 95\\
95\_3,95\_5 & 0.65 & $1.3 \pm 1.59$ & $-1.40 \pm 0.62$ & 1.09  & 95\\ 
95\_3,95\_6 & 1.29 & $0.39 \pm 0.87$ & $-0.40 \pm 0.216$ & 1.08  & 95\\ 
95\_4,95\_5 & -1.03 & $2.14 \pm 2.00$ & $-2.46 \pm 1.19$ & 1.48  & 95\\
95\_4,95\_6 & 2.26 & $0.39 \pm 0.88$ & $-0.40 \pm 0.18$ & 1.05 & 95\\
95\_5,95\_6 & -0.10 & $0.232 \pm 0.68$ & $-0.24 \pm 0.13$ & 1.08 & 95

\end{tabular}
\caption{
The probability enhancement factor is symmetric under the interchange
of the two datasets but $\chi^2_{w12}$ (defined in Eq. 7.9)
is not so we must specify that the datasets column has the
format dataset 1, dataset 2.  
}
\end{center}
\end{table}

\begin{figure}[bthp]
\plotonecita{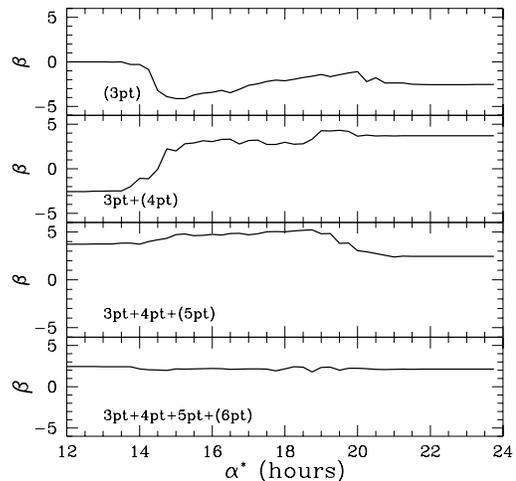}
\plotonecfpa{betapts2_9295_lk.eps}
\caption[betapts2_9295.ps]{\baselineskip=10pt 
The probability enhancement factor $\beta(92,95^*)$, where
the $^*$ indicates that only data with right ascension, $\alpha$,
less than $\alpha^*$ are included.  In the top panel, only the
3 point data are included for 95.  In the panel one lower, 
in addition to all the 3 point data those 4 point 
data with $\alpha < \alpha^*$ are included, etc. 
}
\label{fig:betapts2_9295}
\end{figure}

\begin{figure}[bthp]
\plotonecita{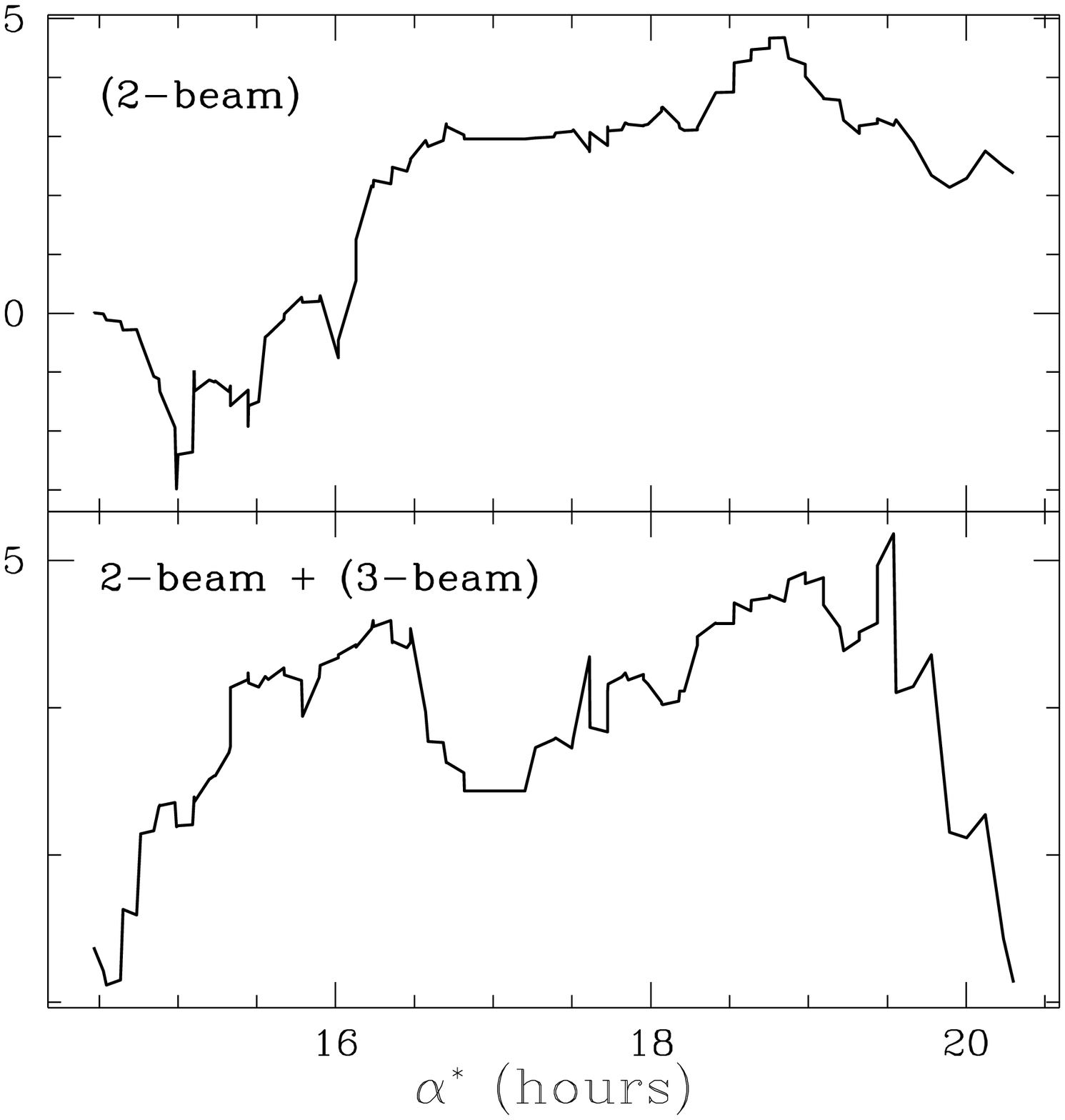}
\plotonecfpa{betapts2_9592.eps}
\caption[betapts2_9592.ps]{\baselineskip=10pt 
The probability enhancement factor $\beta(95,92^*)$.
the $^*$ indicates that only data with right ascension, $\alpha$,
less than $\alpha^*$ are included.  In the top panel, only the
2-beam data are included for 92.  In the bottom panel,
in addition to all the 2-beam data the 3-beam
data with $\alpha < \alpha^*$ are included. 
}
\label{fig:betapts2_9592}
\end{figure}

\begin{figure}[bthp]
\plotonecita{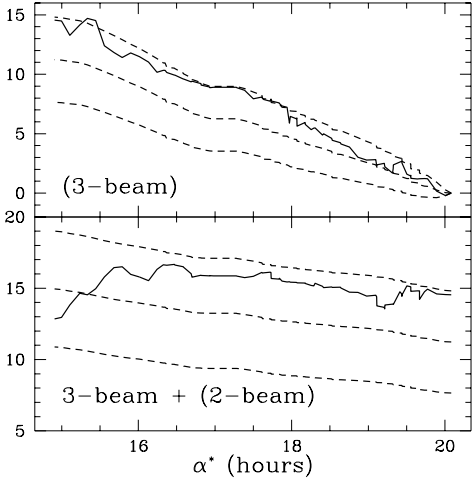}
\plotonecfpa{betapts2_9294_backwards.eps}
\caption[betapts2_9294]{\baselineskip=10pt 
The probability enhancement factor, $\beta(92,94^*)$.
where here $^*$ indicates that only data with right ascension, $\alpha$,
{\it greater} than $\alpha^*$ are included.  In the top panel, only the
3-beam data are being included for 94.  In the
bottom panel, all the 3-beam data are included, but
only those 2-beam data points with $\alpha > \alpha^*$.
The dashed lines are $\langle \beta(92,94^*) \rangle_0$ and 
one standard deviation above and below.} 
\label{fig:betapts2_9294}
\end{figure}

We have also found another breakup of the data to be useful.  To
identify localized problems in the data, we have calculated $\beta$ as
a function of the amount of data included.  For example, in
Fig. \ref{fig:betapts2_9295}, we have plotted $\beta(92,95^*)$
vs. $\alpha^*$, where the star in $95^*$ indicates that only 95 data
with RA $\alpha < \alpha^*$ have been included.  One can see here
features associated with the discrepancies seen in the Wiener filter
figures.  Figures \ref{fig:betapts2_9592} and \ref{fig:betapts2_9294} show
the results of similar calculations.  For Fig. \ref{fig:betapts2_9294},
the order in which the data is included is reversed (see caption) 
in order not to overemphasize the discrepant data at low RA.  

From the first two entries of the Table and also from 
Fig.\ \ref{fig:betapts2_9294} we see that the 3-beam datasets are more
consistent with each other than the 2-beam datasets, where there is a
hint of a problem at low RA.  This hint can also be seen in the
Wiener-filtered data shown in Fig. \ref{fig:f92_shade9294}.  Possibly
confusing is that in Fig. \ref{fig:f92_shade9294} the discrepancy
looks stronger in the 3-beam Wiener-filtered data.  However, this is
because the 2-beam data has significant influence on the best estimate
of the 3-beam signal.  Evidence for this relevance of the 2-beam data
for the 3-beam data (and vice-versa) comes from the fact that
$\beta(92\_2,92\_3)$ and $\beta(94\_2,94\_3)$ are large at $8.3$ and $6.8$
respectively.  A further clue that the problem is with the 2-beam data
is in Fig. \ref{fig:betapts2_9592} where there is a suggestion of a
problem at low RA with the 2-beam but not the 3-beam.

The better agreement between the 3-beam datasets than between the
2-beam datasets is possibly due to the greater susceptibility of
the 2-beam data to atmospheric contamination.  In particular,
the 2-beam data is susceptible to atmospheric gradients while
the 3-beam is not.  A gradient can arise as the 
pendulating motion of the gondola causes the motion of the
chopping flat to be slightly different from constant elevation
\cite{SMpriv}.  Presumably one could test this hypothesis
by searching for signals in the time stream with the balloon pendulation
period.

Both from the Wiener filter figures and the cumulative $\beta$
plot (Fig. \ref{fig:betapts2_9294}) we can see that the MSAM92
and MSAM94 data agree very well at large RA and therefore what's
observed is really on the sky and not some instrumental artifact.
In contrast, the MSAM92/SK Wiener filter figures and
cumulative $\beta$ plots show discrepancies.  These may be due to 
instrumental problems---in which case the problem must be with
SK95---or foreground contamination which could affect either instrument.
We discuss the possibility of foreground contamination in section
\ref{sec:dust} on dust.

\section{Fixing Calibrations by Comparing Datasets}
\label{sec:calib}

Every dataset must be calibrated by using the same
apparatus to observe a radiation source of known brightness.  
This observation allows for the conversion of the
data from some arbitrary units to temperature or brightness units.
Often the brightness of the source in the passband of interest
is only known to 10\% or so in which case the calibration 
is a significant source of uncertainty.  If $\Delta^\prime$
is the uncalibrated data then we define the calibration factor
$f$ as $\Delta = f \Delta^\prime$, where $\Delta$ is the calibrated data.
Similarly, the uncalibrated noise covariance matrix gets adjusted
by $f^2$: $C_n = f^2 C_n^\prime$, since the noise is determined from the
data itself.

One can do likelihood analysis on the uncalibrated
data, but with the appropriate covariance matrix:
\be
C^\prime \equiv \langle \Delta^\prime (\Delta^\prime)^\dagger\rangle =
\langle (s/f) (s/f)^\dagger \rangle + C_n^\prime = 
\left({\sigma_8 \over f}\right)^2\tilde C_T +C_n^\prime.
\ee
For a joint dataset, $\Delta_1$ and $\Delta_2$, we have (dropping the
primes):
\[C =\left( \begin{array}{c}
        \begin{array}{cc}
        \left({\sigma_8 \over f_1}\right)^2\tilde C_{T11} + C_{n11}  &  
         \left({\sigma_8 \over f_1}\right)\left({\sigma_8 \over f_2}\right)
         \tilde C_{T12}    \\ 
        \left({\sigma_8 \over f_1}\right)\left({\sigma_8 \over f_2}\right)
         \tilde C_{T21}   & 
        \ \ \ \ \left({\sigma_8 \over f_2}\right)^2\tilde C_{T22} + C_{n22}
        \end{array} \\
        \end{array} \right) \]

Note that this covariance matrix, and hence the likelihood, 
depends only on $\sigma_8/f_1$ and
$\sigma_8/f_2$.  The degeneracy among the three parameters  
is broken by the calibration measurements of each experiment, which
are usually modeled by a Gaussian:
\be
\ln \L_{\rm tot}\left(\sigma_8,f_1,f_2\right) = \ln \L\left({\sigma_8 \over f_1},{\sigma_8 \over f_2}\right) - {(f_1-\bar f_1)^2 \over 2\sigma_1^2} - {(f_2-\bar f_2)^2 \over 2\sigma_2^2}.
\ee
If one is exploring this likelihood space by direct evaluation, note
that one can first evaluate $\ln \L$ on a two dimensional grid
($\sigma_8/f_1$, $\sigma_8/f_2$) and then evaluate the three-dimensional
$\ln \L_{\rm tot}$ by adding in the (very easy to calculate) calibration
measurement terms.  The data as we receive it has already been
calibrated so we usually take $\bar f = 1$.

We have evaluated this $\ln\L_{\rm tot}$ for the MSAM92 dataset and a
subset of the SK95 ring dataset, from $13^h$ to $22^h$, covering
the range of influence of the MSAM92 data.
For MSAM92 we take $\bar f_{\rm MSAM}= 1$ and $\sigma_{\rm MSAM}=0.1$.
For SK we use the Netterfield et al. calibration $\bar f_{\rm SK}= 1$
and $\sigma_{\rm SK}=0.14$.  We find in this case that $\ln\L_{\rm
tot}$ is minimized at $f_{\rm MSAM}=0.99$, $f_{\rm SK}= 0.99$ and
$\sigma_8=1.13$.  If the Leitch recalibration is used ($\bar f_{\rm
SK}= 1.05$, $\sigma_{\rm SK}=0.07$) then $\ln\L_{\rm tot}$ is
minimized at $f_{\rm MSAM}=1.02$, $f_{\rm SK}= 1.02$ and
$\sigma_8=1.13$.  

We can also neglect the calibration measurements entirely and
use the two datasets themselves to find the best relative calibration,
$f_{12} \equiv f_1 / f_2$.
\be
\L_{12}(f_{12}) \propto \int dx_2 \L(x_1,x_2)\delta(x_1-f_{12}x_2)
\ee
where $x_1 = f_1/\sigma_8$ and $x_2=f_2/\sigma_8$.  We find that,
once again restricting the SK95 data to between 13 and 22 hours
that $f_{\rm MSAM,SK} = 1.06^{+.22}_{-.26}$.  Netterfield et al.
\cite{nett95} find from their analysis that 
$f_{\rm MSAM,SK} = 1.22 \pm 0.24$.

Note that there is a possible problem for joint power spectrum
analysis if relative calibration uncertainty is not taken into
account.  For overlapping experiments neglect of this uncertainty
could artificially boost high frequency power.

\section{Dust}
\label{sec:dust}

\begin{figure}[bthp]
\plotonecita{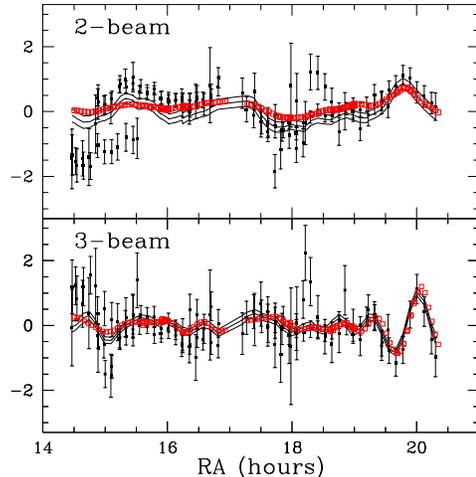}
\plotonecfpa{f92_92_dust_lek.eps}
\caption[f92_92_dust]{\baselineskip=10pt 
Wiener-filtered dust.  The points
with error bars are the MSAM92 pixelized dust data.
Two-beam in top panel, 3-beam
in bottom panel.  The three curves in each panel
are the Wiener-filtered data bounded by $\pm$ \
one standard deviation (assuming Gaussianity!).  
The open squares are the
results of convolving the IRAS data with the MSAM 
beam pattern; the scale is set by a fit to the MSAM data.}
\label{fig:f92_92_dust}
\end{figure}

There is a marginally significant discrepancy between the MSAM
observations and those of Saskatoon at large RA.  This discrepancy is
possibly due to foreground contamination of either the SK or MSAM
datasets.  This hypothesis is supported by the fact that the
discrepancy occurs where the observations are closest to the
plane of the galaxy.  Further, from the MSAM interstellar dust data,
one can see that the discrepancy occurs roughly where the dust is
brightest---see Fig. \ref{fig:f92_92_dust}.

Besides different spectral dependence from the CMB, interstellar dust
also has a different spatial frequency dependence than the CMB.
Schlegel, Finkbeiner, and Davis \cite{DavisFink} have used the DIRBE
and IRAS maps to infer the power spectrum of interstellar dust. They
find that away from the plane of the galaxy it has the shape $C_\ell
\propto \ell^{-2.5}$.  We have therefore used a power spectrum with
this shape to Wiener filter the MSAM dust data---see
Fig. \ref{fig:f92_92_dust}.  The dust is also known to be highly
non-Gaussian.  While the mean signal does not depend on the statistics
of the signal, the uncertainties in the signal do.  Therefore one
should bear this in mind when interpreting the graph since the error
bars in the figure were calculated assuming Gaussianity.  Along with
the MSAM dust data is the result of convolving the MSAM beam pattern
with the IRAS SISSA 240 micron map \cite{Puchalla}.  The IRAS data
have been scaled to fit the MSAM data.  Note that the agreement for
the 3-beam data is much better than for the 2-beam data, once again
suggesting that it is a more reliable dataset.

What we have referred to as the MSAM CMB and dust data are obtained by
fitting each pixel of the four frequency channels (170GHz, 220GHz,
500GHz, 680GHz) of MSAM data to a CMB component and a dust component.
From this fit we get the CMB temperature and dust optical depth.  The
dust is assumed to be a ``grey'' body with temperature $T=20K$ and
emissivity index $\alpha = 1.5$.

Using this model, the dust feature at large RA should not be showing
up in the lowest frequency channel.  Therefore the fit ascribes the
structure seen in the 170GHz channel to CMB.  However, the model may
be an inadequate description; there may be a component correlated
with the dust with stronger emission at 170GHz than the thermal dust
emission itself.  Indeed, the shape of the dust feature at large RA
is somewhat similar to the MSAM CMB feature at large RA.

The Saskatoon data is single frequency and thus harder to directly
check for contamination.  Despite the low frequency, dust (or a source
correlated with dust) contamination is a possibility.  Several
datasets point to a correlation between high frequency maps dominated
by thermal emission from dust and lower frequency measurements
\cite{dustcorr}.  A weak, but significant, correlation has been seen
\cite{Jaffedust} in a correlation analysis of the entire SK dataset
and dust maps made by \cite{DavisFink}; also see \cite{Angelica}.  The
cause of this correlation is not yet known although an hypothesis has
been advanced by Draine and Lazarian \cite{DraineLaz} that it is due
to dipole emission from spinning dust grains.

To investigate this possibility, we have Wiener-filtered the
MSAM92 dust measurements onto the SK95 data in the region of
overlap with MSAM.  For the shape of the dust power spectrum we
used $C_\ell \propto \ell^{-2.5}$ \cite{DavisFink} with amplitude
chosen to maximize the likelihood given the MSAM dust data.  
Although, as expected, the dust is brightest in the region
of the discrepancy, we have been unable to identify any more
detailed relation between the predicted dust signal and the discrepancy.  

\begin{figure}[bthp]
\plotonecita{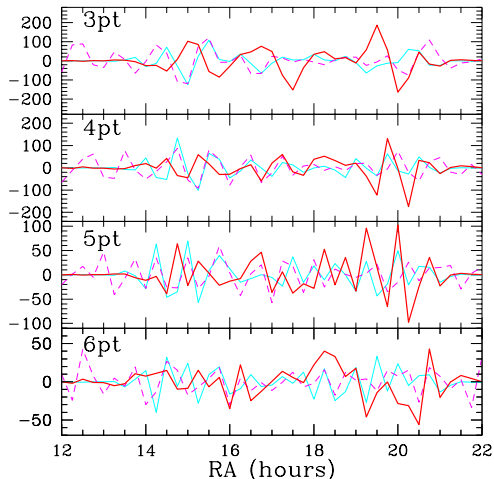}
\plotonecfpa{f95_92,95,92dust_lek.eps}
\caption[f95_92_dust]{\baselineskip=10pt 
The MSAM92 dust data (heavy solid line),
MSAM92 CMB data (light solid line) and SK95 data all
Wiener-filtered on to the SK95 pixels.}
\label{f95_92_dust}
\end{figure}

There is another reason for believing the problem may lie with the SK
data.  The SK team also did some internal consistency checks on their
data, one of which is the A-B test.  Their A and B detectors measured
orthogonal linear polarizations, and thus for an unpolarized, or
weakly polarized, signal, A-B should be consistent with zero.
However, for the region of overlap with MSAM, they find
$\chi^2_{A-B}/\nu = 1.55$ for 80 degrees of freedom.  The origin of
this asymmetry is unclear, possibly an instrumental artifact.  It is
probably too large to be explained by rotational emission from dust
grains since Draine and Lazarian predict that this component of the
dust emission is only between 0.1\% and 10\% polarized.

\section{Summary}

We have demonstrated the usefulness of the Wiener filter for making
visual comparisons of datasets.  We have emphasized that meaningful
consistency testing requires alternative models with which to compare.
Thus we have explicitly extended our model of the data to include a
possible contaminant and calculated the probability distribution of the
amplitude of this contaminant.  For purposes of extracting just one
number from the comparison we advocate calculating the ratio of the
probability of no contamination to the probability of infinite
contamination.  Viewed as a statistic, we have shown this
``probability enhancement factor'' to be better than various $\chi^2$
statistics at discriminating between competing hypotheses.

The utility of our comparison statistics was shown by exercising them on the 
MSAM92, MSAM94 and SK95 data.  
We have found from comparing MSAM92 and MSAM94 that the most probable
level of contamination is 12\%, with zero contamination only 1.05 times
less probable, and total contamination over $2\times 10^5$ times less
probable.  From comparing MSAM92 and SK95 we have found that the most
probable level of contamination is 50\%, with zero contamination only
1.6 times less probable, and total contamination 13 times less
probable.  Looking at subsets of the data we find a region at large RA
where the SK and MSAM measurements disagree.  From IRAS and from the
MSAM dust measurements we know that this region is also the dustiest
region of the overlap between SK and MSAM.  The origin of the
discrepancy is unclear and may be due to instrumental artifacts in SK,
or foreground contamination of either the SK or MSAM measurements.
  
A revolution is underway in the quality and quantity of CMB
data---a revolution generated by the satellites MAP and Planck
\cite{satellites} as well as by a number of balloon and ground-based
programs.  The amount of data may soon be too large for the type of
complete statistical analysis described here.  However, any
approximate methods developed for extracting the power spectrum or
parameters will also be applicable to the statistical procedures
introduced here.

\acknowledgments We thank the MSAM and SK teams for making
their datasets available.  In particular, we are grateful for
assistance from Casey Inman, Jason Puchalla and Grant Wilson.

\end{document}